\documentclass[fleqn,usenatbib]{mnras}

\usepackage{newtxtext,newtxmath}

\usepackage[T1]{fontenc}
\usepackage{ae,aecompl}

\usepackage{graphicx}	
\usepackage{amsmath}	
\usepackage[para,online,flushleft]{threeparttable}

\title{HI content of massive red spiral galaxies observed by FAST}

\author[Lan Wang et al.]{
Lan Wang,$^{1,2}$\thanks{E-mail: wanglan@bao.ac.cn}
Zheng Zheng,$^{3}\thanks{E-mail: zz@bao.ac.cn}$
Cai-Na Hao,$^{4}$
Rui Guo,$^{4}$
Ran Li,$^{1,2}$
Lei Qian,$^{3}$ \and
Lizhi Xie,$^{4}$ 
Yong Shi,$^{5,6}$
Hu Zou,$^3$
Yixian Cao,$^7$
Yanmei Chen,$^{5,6}$
Xiaoyang Xia$^{4}$
\\
  $^1$Key Laboratory for Computational Astrophysics, National
Astronomical Observatory, Chinese Academy of Sciences,\\
Datun Road 20A, Beijing 100012, China \\
$^{2}$School of Astronomy and Space Science, University of Chinese Academy of Sciences, Beijing 100049, China \\
$^3$National Astronomical Observatories, Chinese Academy of Sciences, Datun Road 20A, Beijing 100101, China \\
  $^4$Tianjin Astrophysics Center, Tianjin Normal University, Tianjin 300387, China \\
  $^5$School of Astronomy and Space Science, Nanjing University, Nanjing 210093, China \\
  $^6$Key Laboratory of Modern Astronomy and Astrophysics (Nanjing University), Ministry of Education, Nanjing 210093, China \\
  $^7$Max Planck Institute for Extraterrestrial Physics, Giessenbachstrasse 1, 85748 Garching Germany
}

\date{Accepted XXX. Received YYY; in original form ZZZ}

\pubyear{2022}

\begin{document}
\label{firstpage}
\pagerange{\pageref{firstpage}--\pageref{lastpage}}
\maketitle

\begin{abstract}
A sample of 279 massive red spirals was selected optically by Guo et al. (2020), among which 166 galaxies have been observed by the ALFALFA survey. In this work, we observe HI content of the rest 113 massive red spiral galaxies using the Five-hundred-meter Aperture Spherical radio Telescope (FAST). 75 of the 113 galaxies have HI detection with a signal-to-noise ratio (S/N) greater than 4.7. Compared with the red spirals in the same sample that have been observed by the ALFALFA survey, galaxies observed by FAST have on average a higher S/N, and reach to a lower HI mass. To investigate why many red spirals contain a significant amount of HI mass, we check color profiles of the massive red spirals using images observed by the DESI Legacy Imaging Surveys. We find that galaxies with HI detection have bluer outer disks than the galaxies without HI detection, for both ALFALFA and FAST samples. For galaxies with HI detection, there exists a clear correlation between galaxy HI mass and g-r color at outer radius:  galaxies with higher HI masses have bluer outer disks. The results indicate that optically selected massive red spirals are not fully quenched, and the HI gas observed in many of the galaxies may exist in their outer blue disks. 

\end{abstract}




\section{Introduction}

It has been well-established that galaxies roughly fall into two groups in the color-mass diagram: one group is dominated by red quiescent galaxies and the other mainly consists of blue star-forming galaxies \citep{baldry2004, baldry2006}. Morphological studies revealed that red sequence galaxies are mostly early-type galaxies and blue galaxies are late-type ones \citep[e.g.][]{schawinski2014}. Theoretical studies proposed several mechanisms that can cease star formation in massive galaxies \citep{faber2007, fang2013, guorui2016a, guorui2016b, xu2020star}, during which morphological transformation from disk-dominated systems to bulge-dominated systems usually accompanies.
However, massive red passive spiral galaxies were found in both clusters and field environments, in the local Universe and at redshift as high as $\sim 1$ \citep{bundy2010, xu2021giant}, indicating that these spiral galaxies can quench their star formation without experiencing morphological transformation.
It is therefore important to investigate more the various  properties of massive red spiral galaxies, to understand their formation and quenching in detail.

Many works have been done to study the optical properties of passive spiral galaxies \citep[e.g.][]{Masters2010, Robaina2012, Fraser-McKelvie2018}.
Recent studies of \citet{guorui2020} and \citet{hao2019} focused on a sample of 279 massive red spiral galaxies optically selected from SDSS DR7, and investigated a subset of these galaxies with MaNGA observations. They found that both the disk and bulge components of massive red spirals formed earlier and in shorter timescales than those of massive blue spirals. Based on the same set of MaNGA data, \citet{zhoushuang2021} reached a similar
conclusion via a Bayesian analysis on the star formation histories. However,
the quenching mechanisms for the massive red spirals are still unclear.

Gas content is the key component to understand galaxy formation and quenching as it is the requisite fuel for star formation. The quenching of massive red spirals is inevitably related to the insufficient supply or collapse of cold gas, regardless of the details of various mechanisms (e.g., AGN feedback, halo gas shock heating, morphological quenching) that stop star formation \citep{tacconi2020ARAA}. Although molecular gas is considered to be more directly linked to star formation \citep[e.g.][]{bigiel2008}, atomic gas is a probe of gas reservoir in general and can serve as the first step to understand the quenching processes that affect gas content. \citet{guorui2020} studied the HI content in red spirals by cross-matching with the ALFALFA catalog.  The cross-identification with ALFALFA led to 166 galaxies, among which 74 have HI detections.

To obtain a complete census of the HI content of the massive red spiral galaxy sample of \citet{guorui2020}, we observe the remaining 113 red spirals that are not covered by ALFALFA using FAST. This almost doubles the sample size of massive red spirals with HI observations and hence will enable us to have a better statistics of the HI content of such systems. 
In addition, the HI observations obtained by FAST can be used to compare FAST with Arecibo in their capabilities in detecting atomic gas of galaxies.

\citet{guorui2020} found that $\sim 45\% (74/166)$ of the red spirals within the ALFALFA coverage have HI detections, similar to the blue spirals. While these galaxies have overall red color, it is interesting to find out whether the atomic gas is related to any residual star formation and where the HI content reside in these galaxies. Unfortunately single dish observations obtained by FAST and Arecibo only provide total amount of HI fluxes in galaxies, and spatially resolved HI observations that can locate in detail the origin of atomic gas are very expensive and have been applied only to small samples \citep[e.g.][]{lemonias2014}. 
Optical blue color normally indicates the existence of young stellar populations and possibly recent star formation activities. Thus a blue component may also indicate the existence of HI content. In this work, we examine galaxy color profile in the massive red spirals, to get a clue of the location of HI content in these galaxies indirectly.

The paper is organised as follows. In section 2, we first introduce briefly the sample of massive red spiral galaxies we use, then describe in detail how we analyse the spectra observed by FAST to get HI content of galaxies. In section 3, we first show FAST observation of 6 test galaxies that have been observed by ALFALFA, and compare the detailed spectra and the derived HI masses in the two observations.
Then we present results of HI detection for 113 massive red spiral galaxies observed by FAST, and compare the statistics with the galaxies covered by ALFALFA. In section 4, we study the color profiles of the massive red spirals, and show that galaxies with larger HI mass have in general bluer outer disks, where the HI gas should reside in. Conclusions and discussions are presented in the final section.

\section{sample and observation}

\subsection{massive red spirals}
In \citet{guorui2020}, a sample of face-on massive red spiral galaxies are selected based on the SDSS DR7 catalogue of \citet{mendel2014}. The selection criteria include:
\begin{itemize}
\item[-] $0.02<z<0.05$, and total stellar mass greater than $10^{10.5}M_{\odot}$
\item[-] visual morphological type to be spirals, with minor-to-major axis ratio b/a$\ge 0.5$ 
\item[-] red sequence galaxies in the dust-corrected u-r color -- stellarmass diagram
\end{itemize}

\begin{figure}
\hspace{-0.4cm}
\resizebox{8.5cm}{!}{\includegraphics{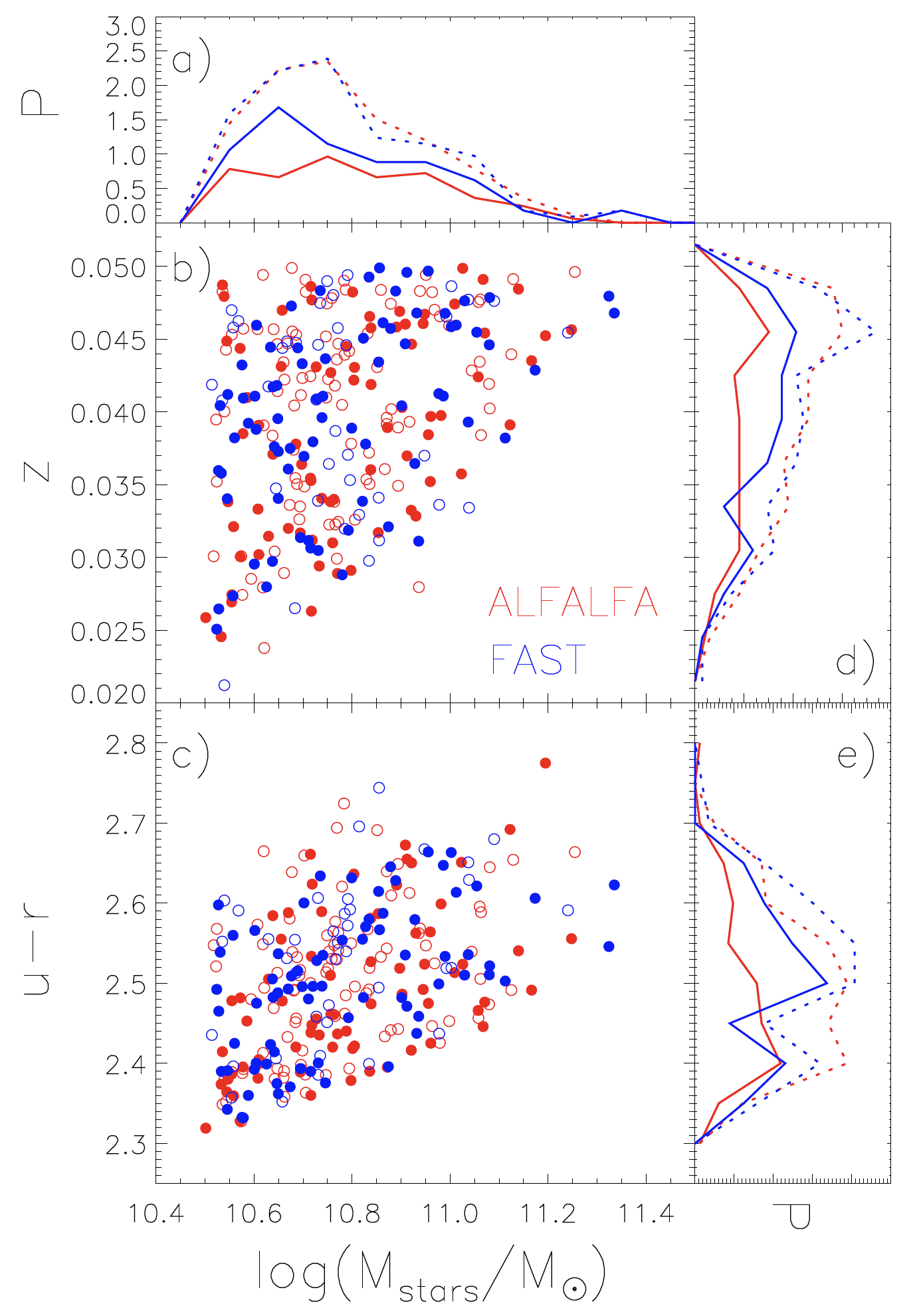}}\\
\caption{
Stellar mass, redshift and dust-corrected u-r color distributions of the massive red spirals in the ALFALFA region (red symbols) and in this work observed by FAST(blue symbols). In panels b) and c), open circles are for galaxies without HI detection in either observation, and solid circles are for galaxies with HI detection (with S/N greater than 4.7, see section 2.2 for the details). In panels a), d) and e), distributions of stellar mass, redshift and dust-corrected u-r color are shown respectively for the galaxies in the ALFALFA region (red lines) and observed by FAST(blue lines). Dotted lines are distributions for all the galaxies in each subsample, and solid lines are contributions from the galaxies with HI detection.
}
\label{fig:Mstar_z}
\end{figure}

The details can be found in \citet{guorui2020} and references therein. Based on the criteria above, in total 279 massive red spiral galaxies are selected. In panel b) and c) of Fig.~\ref{fig:Mstar_z}, the sample galaxies are plotted in the diagram of stellar mass versus redshift, and the diagram of stellar mass versus dust-corrected u-r color. 166 of these galaxies are in the coverage of ALFALFA survey, and are plotted in red. The rest 113 galaxies are not in the ALFALFA region, and are plotted in blue. In Fig.~\ref{fig:Mstar_z}, we also show the distributions of stellar mass, redshift and color for the sample galaxies in panels a), d) and e). Dotted red (blue) lines are for all the galaxies in ALFALFA (FAST). The stellar mass and redshift distributions are similar in the two subsamples, while the galaxies observed by FAST are in general a bit redder than the galaxies in ALFALFA. In this work, we focus on these 113 galaxies and use FAST to observe their HI content. 

\subsection{HI detection by FAST}
We use the Five-hundred-meter Aperture Spherical radio Telescope \citep[FAST;][]{Nan2006,jiangpeng2019,qian2020} to observe 113 massive red spirals, which are not in the region of ALFALFA survey. We observe an extra of 6 test galaxies that have been observed by ALFALFA, for a comparison between the signals detected by the two telescopes.

FAST, with a diameter of 500 m, has strong power in detecting HI 21cm spectra at cosmological distances. Its 19-beam receiver covers a frequency range of 1.05–1.45GHz. The angular resolution is about 2.9 arcmin. The system temperature (including sky background) is about 20K for observations with zenith angle within 26.4 deg. 

We observed our target galaxies through a ‘Shared-Risk’ project of FAST (project ID: 2019a-133-O). The observations were made on the dates of 20190627, 20191216-20191219, 20200103 using the position switch ON-OFF mode. We record two polarizations (xx and yy) for all targets. For 13 galaxies that were observed in the first round on 20190627, we observe the targets with the central beam (M01) only. The ON-source and OFF-source integration time are set to be 45 seconds, and the switch time between ON-source and OFF-source positions (overhead) is 30 seconds. For all other 100 galaxies, and the 6 test galaxies, we follow the observation strategy of \citet{Zheng2020}, i.e. covering the target using M01 and M14 beams in turns during ON-source and OFF-source mode respectively, so that the M14 (M01) is ON-source when the M01 (M14) is OFF-source. We set a longer integration time, 120 seconds, for both ON-source and OFF-source positions, to be able to detect weaker HI signals in the galaxies. Therefore, the total ON-source integration time for these galaxies is 240 seconds (M01 + M14). We inject the high (10K) noise diode signal for 1s every 5s during the observation. 

\begin{figure}
\hspace{-0.4cm}
\resizebox{8cm}{!}{\includegraphics{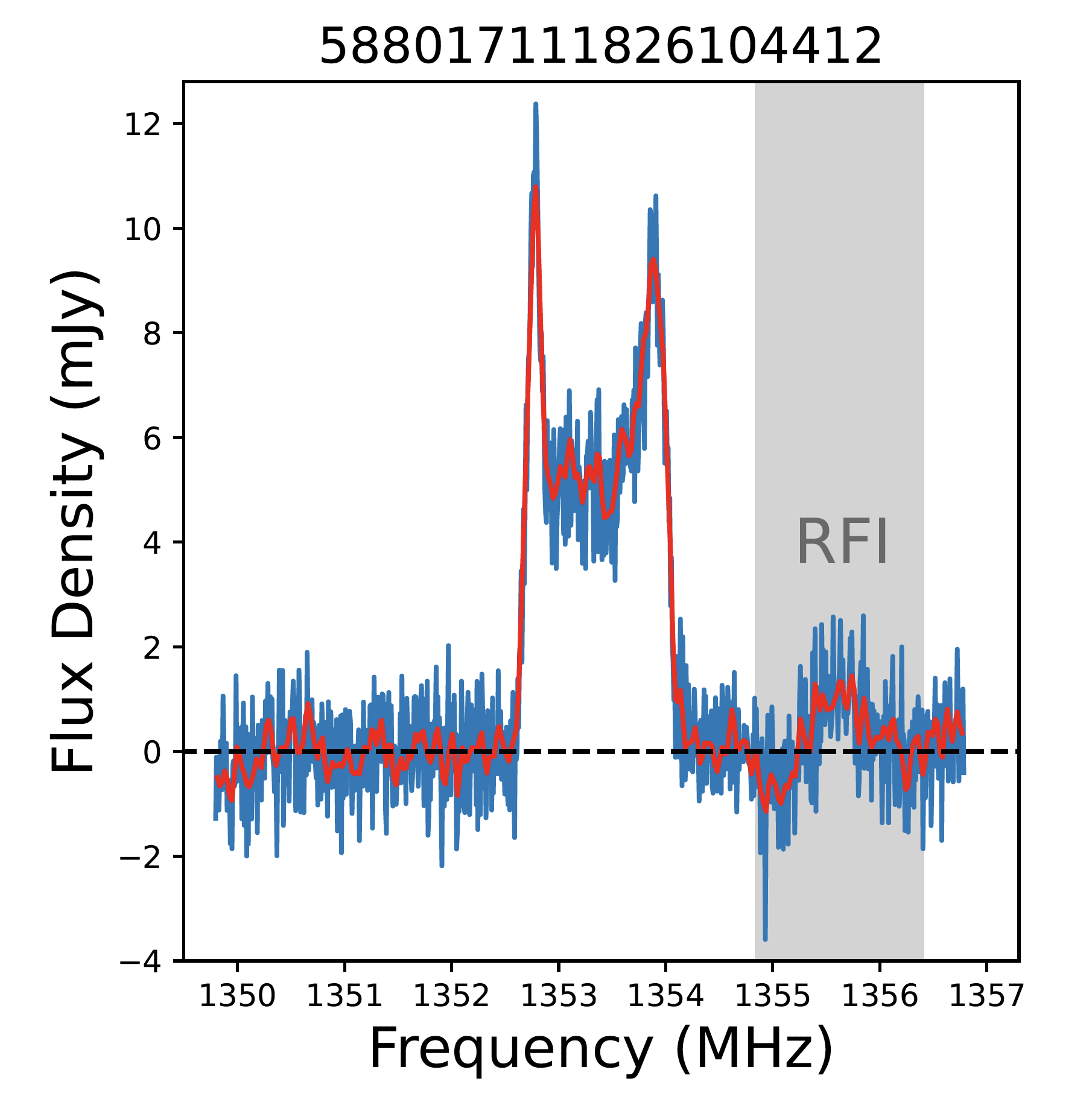}}\\
\caption{
An example of the HI 21cm spectrum of a galaxy in our sample. The blue line shows the original spectrum and the red line shows the spectrum smoothed to a resolution of $\sim 10 \rm km s^{-1}$ for analysis. An RFI is indicated in the shaded region. The horizontal dashed line marks zero flux as a reference.
}
\label{fig:example_FAST}
\end{figure}

We list our data reduction and analysis procedures below:

\begin{itemize}

\item[-] Flux calibration. We use the noise diode signal to calibrate the recorded power into antenna temperature in unit of K. We then transform the temperature into flux density in unit of Jy using the gain values listed in \citet[Table 5 of ][]{jiangpeng2020}, for M01 and M14 beams respectively. 

\item[-] Ripple removal. The FAST spectra are contaminated by $\sim 1$ MHz ripples caused by reflections between the feed cabin and the reflecting surface. Following \citet{Zheng2020}, we remove the 1 MHz ripple by subtracting a sinusoidal function fitted using spectra around the HI frequency of the target. 

\item[-] Continuum subtraction. Two continuum frequency regions without RFI contamination at both sides of the HI signal are selected by eye. We fit a straight line using the two regions to get the continuum and subtract it from the spectrum to get the HI spectrum. 

\item[-] Spectra smoothing. Our raw data is recorded with a frequency resolution of $7.6$ kHz, corresponding to a velocity resolution of $\sim 1.7 \, \rm km s^{-1}$ at 1420 MHz. The ALFALFA spectra have a lower velocity resolution, $\sim 5.5 \, \rm km s^{-1}$, which have been Hanning smoothed to  $\sim 10 \, \rm km s^{-1}$ to be analysed as shown in \citet{haynes2018}. In order to compare with the results of ALFALFA, we also smooth our spectra to a resolution of $\sim 10\, \rm km s^{-1}$. First, we combine every three FAST frequency channels into one, and set the flux density value to be the mean of the three channels. This gives us a velocity resolution of  $\sim 5 \, \rm km s^{-1}$. Then we apply Hanning smooth to the combined data points, and get a resolution of  $\sim 10 \, \rm km s^{-1}$. The rms noise of the spectrum $\sigma_{\rm rms}$ (in unit of mJy) is calculated on the two continuum regions at the two sides of the HI signal, after continuum subtraction.

\item[-] Derivation of $W_{50}$, the velocity width of the HI profile at 50\% level of the profile peaks. 
Two observed frequencies $\nu_1$ and $\nu_2$ that correspond to half of the maximum flux density values on each horn of the line are selected, and $W_{50}$ is defined as the difference between the two velocities that correspond to the two frequencies. In practice, as in \citet{haynes2018}, the maximum flux density value is calculated as the observed peak flux density minus the rms, to correct for the contribution of noise. Following equation (20) of \citet{meyer2017}, we calculate the rest frame velocity width as:
\begin{equation}
W_{\rm 50} = \frac{\nu_2 - \nu_1}{\nu_0}c,     
\end{equation}
where $\nu_0$ is the observed central HI frequency of the target galaxy, and $c$ is the speed of light. Note that this is a factor of $1+z$ narrower than the observed frame values calculated and presented in \citet{haynes2018}, as indicated in euqation (21) of \citet{meyer2017}.

\item[-] Calculation of the rest frame velocity integrated HI line flux, $S_{21}$ (in unit of $\rm Jy\,km s^{-1}$). The central frequency of the HI line is determined by the optical spectroscopic redshift of the galaxy and the frequency range of the HI line is selected by hand. The total HI flux is then calculated by integrating the continuum subtracted HI flux density over this frequency range. 
\item[-] Estimation of the S/N of the detections. To be compared with the S/N provided by ALFALFA catalogue, following \citet{haynes2018}, the S/N of the detection is estimated as:
\begin{equation}
    {\rm S/N} = (1000\, S_{21} /W_{50})\times w_{\rm smo}^{1/2} / \sigma_{\rm rms},
\end{equation}
$w_{\rm smo}$ is a smoothing width, which is basically the $W_{50}/2$ in unit of $10 \rm km/s$. In reality, $w_{\rm smo}$ is set to be $W_{50}/20$ for $W_{50}<400 \rm km s^{-1}$ and 20 for $W_{50}\ge 400 \rm km s^{-1}$. The uncertainty of the integrated line flux $\sigma_{S_{21}}$ is calculated as $S_{21}/(\rm S/N)$. 

\item[-] Derivation of the HI mass and uncertainty. The HI mass of the galaxy is derived from the integrated HI flux and distance following the equation provided by \citet{meyer2017}, under the assumption that the HI content is optically thin:
\begin{equation}
    \frac{M_{\rm HI}}{M_{\odot}}=\frac{2.35\times 10^5}{1+z}\times \frac{S_{21}}{\rm Jy\,km s^{-1}} \times \frac{D}{\rm Mpc}^2.
\end{equation}
$D$ is the luminosity distance of the galaxy in unit of Mpc calculated from the spectroscopic redshift of the galaxy.
The error of the HI mass is calculated similar as adopted by \citet{haynes2018} to include systematic uncertainty in the flux calibration, where the uncertainty of distance has been ignored:
\begin{equation}
    \sigma_{\rm log M_{\rm HI}} = \sqrt{(\sigma_{S_{21}}/S_{21})^2+0.1^2} /\ln10,
\end{equation}
\end{itemize}

Fig.~\ref{fig:example_FAST} gives an example of our observed spectrum of one galaxy with HI detection, after processed following the above steps. The spectrum of this galaxy of z=0.0496 contains a typical HI line with two horns. 
An RFI signal is indicated in the shaded region, which was generated by the refrigerating dewar in the compressor \citep{jiangpeng2020, Xi2022}\footnote{The compressor has been screened in Aug 2021, and the contamination of this type of RFIs has been removed in observations afterwards.}.  
$W_{50}$ of this HI line is 294 $\rm km s^{-1}$, $\sigma_{\rm rms}$ = 0.34 mJy, S/N = 76.5, $S_{21}$= 1.97 Jy $\rm km s^{-1}$, and log$M_{\rm HI}$=10.33, with an error of 0.04.

Note that the 13 galaxies observed in the first round have shorter integration time than the rest 100 galaxies, and therefore have in general larger $\sigma_{\rm rms}$. The median $\sigma_{\rm rms}$ for the 13 galaxies is 0.79 mJy, and for the rest 100 galaxies the median $\sigma_{\rm rms}$ is 0.39 mJy.
2 of the 13 galaxies have S/N$>6.5$, and the rest 11 have no HI detection. We have checked that when excluding the 13 galaxies, the results shown in the following sections remain qualitatively similar. For completeness, we include all 113 galaxies observed by FAST in the analysis.

We also note that for the 166 galaxies that have been observed in ALFALFA, the integration time is $\sim$ 48 seconds. For the massive red spirals with HI detection in ALFALFA, the median rms noise is 2.35 mJy. This number is larger than the median rms noise of 0.79 mJy for the 13 galaxies observed by FAST in the first round, which have integration time of 45 seconds. With similar integration time, the difference in rms noise reflects different performance of the two telescopes used.

The angular resolution of 2.9 arcmin for FAST corresponds to $\sim 70$kpc at z=0.02 and $\sim 170$kpc at z=0.05. In a few cases, our target galaxies have close companions, and the observed value may include flux from both the target galaxy and its close neighbor. 

\begin{figure*}
\hspace{-0.4cm}
\resizebox{17.cm}{!}{\includegraphics{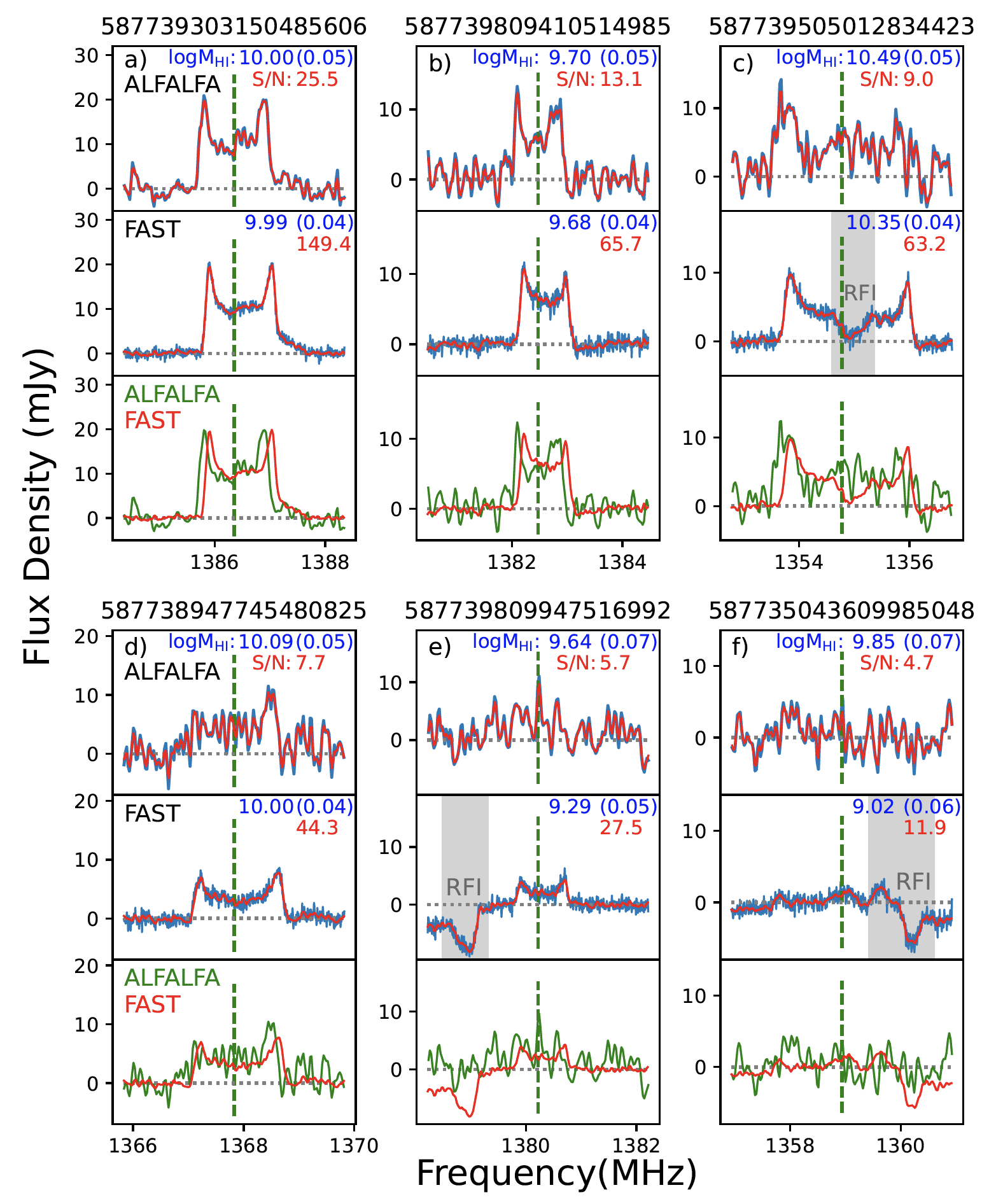}}\\
\caption{
Spectra of the 6 test galaxies observed both in ALFALFA and by FAST. For each galaxy, its SDSS object ID is listed on top of the three panels, with the upper panel showing result of ALFALFA, the middle panel showing result of FAST and the bottom panel comparing the two. In the upper and middle panel, the grey-blue line shows the original spectrum and the red line shows the spectrum smoothed to a resolution of $\sim 10 \rm km s^{-1}$. The signal-to-noise ratios are listed by red numbers, and derived HI masses are listed in blue with errors in parentheses. The bottom panel for each galaxy compares the smoothed spectra of ALFALFA (green line, identical to the red line in the upper panel) and FAST (red line, identical to the red line in the middle panel). Green vertical dashed lines indicate the central frequency calculated from the spectroscopic redshift of the galaxy. Galaxies are listed in order of high to low signal-to-noise ratios from galaxy a) to galaxy f). RFIs are indicated in the shaded regions for the FAST spectra.
}
\label{fig:test_ALFA}
\end{figure*}

\begin{figure}
\hspace{-0.4cm}
\resizebox{8cm}{!}{\includegraphics{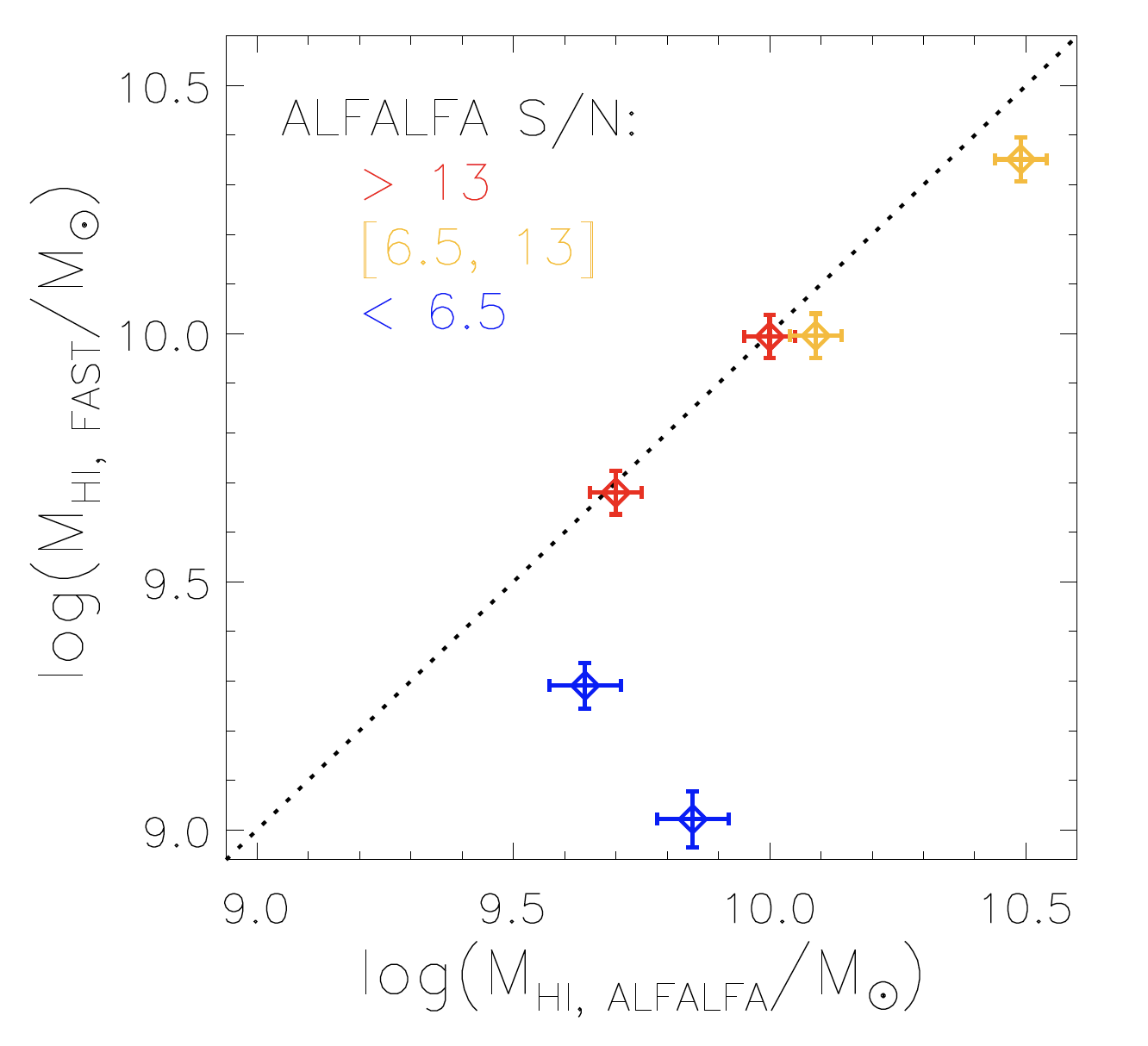}}\\
\caption{
For the 6 galaxies that have both observed by ALFALFA and FAST, comparison of the HI mass detected in the two observations. Red symbols are for the 2 galaxies with S/N > 13 in ALFALFA. Blue symbols are for the 2 galaxies with S/N <6.5, and orange symbols are for the 2 galaxies with intermediate S/N in ALFALFA.}
\label{fig:test_6galaxies}
\end{figure}

\section{Results of HI detection}

Before presenting results of our sample galaxies observed by FAST, we first show results for the 6 test galaxies that have been observed both in ALFALFA and by FAST in this work. We compare the spectrum and HI mass obtained in the two observations, to have an idea of the difference in between. After that, we show results of the HI detection of our 113 massive red spirals, and compare with the ones in the coverage of ALFALFA.

\subsection{HI detection of 6 test galaxies}

In Fig.~\ref{fig:test_ALFA}, for each of the 6 test galaxies, we show and compare their spectra from the two observations, with the upper panel showing results from ALFALFA \citep{haynes2018}, the middle panel from our observation by FAST, and the bottom panel comparing the two. For each galaxy, in the upper and middle panels, the grey-blue line shows the original spectrum and the red line shows the spectrum smoothed to a resolution of $\sim 10 \rm km s^{-1}$. The FAST raw spectra we observe are sampled at $\sim 1.7 \rm km s^{-1}$ at $z \sim 0$, while for ALFALFA the value is $\sim 5.5 \rm km s^{-1}$, with differences clearly seen in the panels. More RFIs show up in the FAST observation than in ALFALFA. Another clear difference is that the rms of noise is much smaller in FAST observation than in ALFALFA, partially due to the longer integration time used for FAST observation. In Fig.~\ref{fig:test_ALFA}, galaxies are listed in an order from highest S/N to lowest S/N, from galaxy a) to galaxy f). As described in Section 2.2, S/N of the HI lines are calculated in a similar way in both observations, and the numbers are listed in red in each panel. Observations by FAST have higher S/N for the same galaxy, due to smaller rms noise. For galaxies with high S/N, the comparison in the bottom panels show a small systematic shift between the spectra of the two observations, probably due to the velocity change of the earth.

Blue numbers listed in each panel of Fig.~\ref{fig:test_ALFA} are the HI mass derived in the corresponding observation, with errors listed in parentheses. For the two galaxies a) and b) that have highest S/N, the HI lines are clearly seen in both observations and have similar shapes. Their derived HI masses are quite consistent. For the following two galaxies c) and d) with smaller S/N, which still have S/N greater than the approximate minimum S/N threshold of 6.5 in ALFALFA \citep{saintonge2007}, the HI lines becomes noisier in the ALFALFA observation, while spectra obtained by FAST still show clear line shapes. The derived HI masses have a bit larger difference in the two observations. For galaxy c), there exists a dip near the central frequency of the galaxy for the FAST observation. We have checked that this is caused by an RFI that happens to overlap with the HI line. The HI flux derived using the FAST data could be underestimated because of the RFI effect. For galaxy d), the two observations give identical W50, but there exists an excess of flux on the right horn of the HI line in ALFALFA, and there exists a weak RFI in the FAST observation which may affect the flux derived.

For the two galaxies e) and f) with S/N smaller than 6.5, the HI lines are weak in the results of ALFALFA, and the differences in HI mass are even larger. 
For galaxy f) which has smallest S/N, we have checked that the large difference of HI mass derived in the two observations is mainly due to the different line widths identified. FAST result of galaxy f) gives W50=106 $\rm km s^{-1}$, which corresponds to a frequency width of 0.48MHz. For ALFALFA result, W50 is 304 $\rm km s^{-1}$ as provided by \citet{haynes2018}, with a line width about three times as the one identified in the spectrum of FAST. For galaxy e), W50 is 273 $\rm km s^{-1}$ (which corresponds to 1.3MHz) in ALFALFA result, much wider than that in the FAST observation. 

Fig.~\ref{fig:test_6galaxies} summarises the mass comparison of the 6 galaxies, where consistency and differences can be seen more directly. From the detailed comparison of spectra as shown in Fig.~\ref{fig:test_ALFA} and discussions above, we find that only galaxy a) and b) have proper measurements from both ALALFA and our FAST project. For these two galaxies with S/N$>6.5$ in ALFALFA (red symbols in Fig.~\ref{fig:test_6galaxies}), the HI masses derived are almost the same in the two observations. For the two galaxies with $6.5<$S/N$<13$ (orange symbols), the HI mass of galaxy c) is underestimated a bit in FAST due to the effect of an RFI, as seen in Fig.~\ref{fig:test_ALFA}, and the HI mass of galaxy d) derived from ALFALFA observation is about 0.1dex larger than that from FAST. For the two galaxies with S/N$<6.5$ in ALFALFA  (blue symbols), FAST observations which have higher S/N give much smaller HI masses, indicating that the HI mass is probably over-estimated in ALFALFA for these low S/N galaxies. The errors plotted in Fig.~\ref{fig:test_6galaxies} and listed in Fig.~\ref{fig:test_ALFA} are calculated as described in \citet{haynes2018} for ALFALFA galaxies and in section 2 for FAST galaxies. 

We should note that for ALFALFA, HI fluxes are obtained based on HI cubes during sky scans, while for our FAST observation, the position switch ON-OFF mode is applied. Therefore, pointing errors may cause some lost of flux in the FAST observation. According to \citet{jiangpeng2020}, pointing errors of the 19-beam receiver of FAST in different sky positions are less than 16 arcsec. The standard deviation of pointing errors is 7.9 arcsec, which is less than one-20th of the beam width near 1450MHz. Assuming that the target is a point source and the FAST beam has a Gaussian shape with a FWHM of 2.9arcmin \citep{jiangpeng2020}, a 16 arcsec offset in the sky would lead to an underestimation of 2 per cent in flux, which corresponds to a 0.01 dex offset. Besides, 
for galaxies that fill the FAST beam partially, due to the approximately Gaussian response on the sky, our FAST observation may lose flux for the outer part of the HI discs.  According to the tight HI mass-size relation for disc galaxies \citep{wangjing2016}, we have checked that most of our galaxies have an HI diameter less than 1 arcmin. Assuming a flat HI profile with a diameter of 1 arcmin, a Gaussian beam shape with FWHM of 2.9 arcmin could recover 96 per cent of the total HI flux, which corresponds to a 0.02 dex offset. Therefore, the total underestimation of the HI flux by FAST due to the above effects should be less than $\sim$ 0.03 dex. 

\subsection{HI mass of massive red spirals}
Table \ref{table:catalog} presents the results of the 113 massive red spirals observed by FAST.
The catalogue of \citet{haynes2018} includes HI detections of high signal-to-noise ratio which have S/N $> 6.5$, as well as detections of lower S/N for the ones that have optical counterparts. For the 166 massive red spirals in the ALFALFA catalogue that are considered to have HI detection, the minimum S/N is 4.7. We therefore set the minimum S/N to be 4.7 for the galaxies observed by FAST, above which we consider the galaxy to be detected with HI content. Table \ref{table:numbers} lists the numbers of massive red spirals in the ALFALFA coverage and observed by FAST in this work, and the numbers of  galaxies with S/N greater than 4.7 and 6.5 respectively. The corresponding detection rates are listed in parentheses. Subsample of galaxies observed by FAST has a higher detection rate, due to in general much lower rms noise and therefore higher S/N of the detection.

\begin{table*}
    \caption{Results of the 113 massive red spirals observed by FAST. 10 galaxies are listed below and the full table is available online. }
    \begin{threeparttable}
    \begin{tabular}{llllllllll}
    \hline
    objID & ra & dec & z  & $W_{50}$($\rm km\,s^{-1}$)  & $\sigma_{\rm rms}$(mJy)    &  $S_{21}$ (Jy\,$\rm km\,s^{-1}$)  & $\rm log(M_{\rm HI}/M_{\odot})$  &  $\sigma_{\rm log M_{\rm HI}}$  & S/N \\
    (1)   &  (2)   &   (3)  &  (4)   &  (5)  &  (6)   &  (7)    &  (8)    &  (9)     & (10)   \\
    \hline
    587725774534148388   &   122.377   &   47.7896  &  0.0408   &    & 0.41   &     &      &      &    \\
    587728679539245209   &   170.779  &    65.2518  &   0.0476   &    253 &  0.40 &    0.70   &   9.85  &  0.05   &   24.9 \\
    587727229986078786   &   26.444  &   -9.7429 &   0.0494   &   10 & 0.46  &  0.03  &    8.44  &   0.12   &   3.9 \\
    587727178981769320   &   349.363   &  -10.0307  &  0.0336   &    & 0.54  &      &      &      &     \\
    587730845817504193   &   321.890   &  -1.1886  &  0.0305   &   298 & 0.42   &   2.61   &   10.03 &   0.04  &    79.6 \\
    587725816408768584   &   158.830   &   64.1775  &  0.0411    &  176 & 0.41   &  0.14    &  9.02   & 0.09  &    5.8 \\
    588009366404726908   &   157.984   &   59.0177  &  0.0458   &     & 0.34    &     &     &      &     \\
    587727180610338874   &   30.955   &  -8.1289   & 0.0413   &   75 & 0.44    & 0.17   &   9.11 &   0.06   &   10.0 \\
    587731511536976010    &  30.033   &  -1.0156  &  0.0404    &  266 & 0.45  &    1.40   &   10.00  &  0.04  &    42.7 \\
    588015507667353745   &   19.634   &  -1.1974  &  0.0468    &  91  & 0.45   &  0.54    &  9.72  &  0.05   &   28.4 \\
    ... \\
    \hline  
    \end{tabular}
    \begin{tablenotes}
        \item[] {\bf Notes.} Columns: (1) SDSS object ID, (2) $\&$ (3) coordinate of the galaxy, (4) spectroscopic redshift of the galaxy, (5) velocity width of the HI profile, (6) rms noise of the spectrum, (7) rest frame velocity integrated HI line flux, (8) HI mass of the detection,  (9) uncertainty of HI mass, (10) S/N of the detection. 
    \end{tablenotes}
    \end{threeparttable}
\label{table:catalog}   
\end{table*}

\begin{figure}
\hspace{-0.4cm}
\resizebox{8cm}{!}{\includegraphics{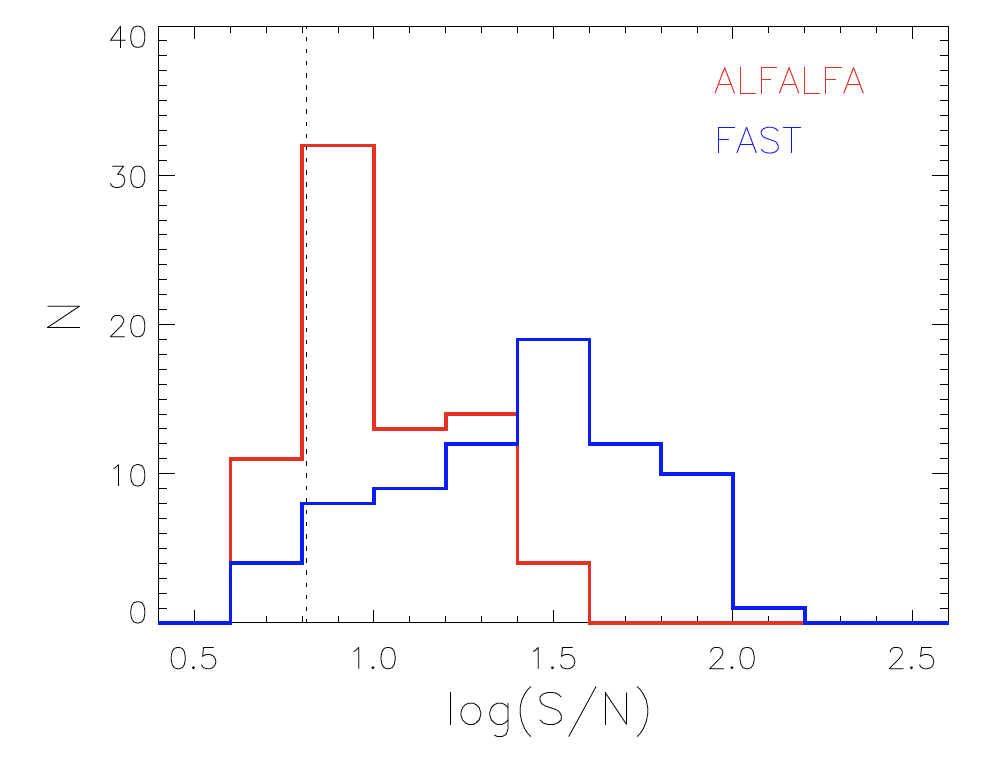}}\\
\caption{
Distributions of the signal-to-noise ratio of the massive red spiral galaxies observed by ALFALFA (red) and FAST (blue), for galaxies with HI detection (with S/N greater than 4.7). Vertical dotted line indicates the threshold of S/N=6.5.
}
\label{fig:HI_SN}
\end{figure}

Fig.~\ref{fig:HI_SN} gives the S/N distributions for the massive red spirals in ALFALFA and observed by FAST, for galaxies with S/N$>4.7$. This figure shows more clearly that the S/N is in general much higher in the FAST observation than in ALFALFA. S/N peaks at around 8 for the ALFALFA galaxies, and peaks at around 32 for the FAST galaxies. The much lower rms noise observed by FAST in this work leads to the fact that weaker signals and hence galaxies with lower HI masses are detected when we set the same threshold of S/N to define galaxies with HI detection. 

\begin{figure}
\hspace{-0.4cm}
\resizebox{8cm}{!}{\includegraphics{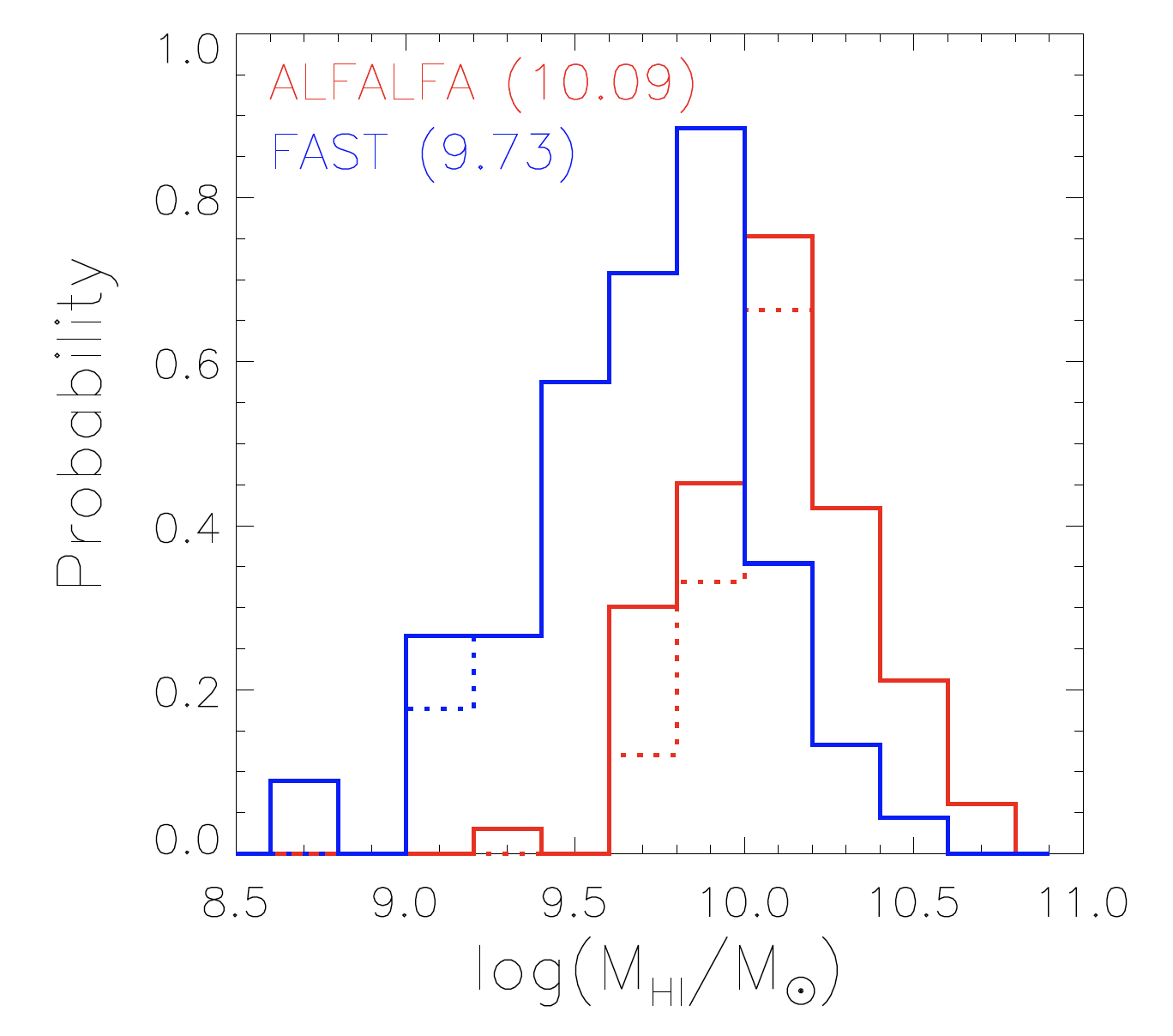}}\\
\caption{
HI mass distributions for the massive red spiral galaxies with HI detection observed by ALFALFA (red) and FAST (blue). The probability is normalised to the total number of galaxies in each subsample. Solid histograms show results of galaxies with $S/N>4.7$. Dotted histograms are the results of galaxies with $S/N>6.5$, which are on top of the solid histograms in most of the mass bins. The number indicated in parentheses is the median logarithmic HI mass of the galaxies with $S/N>4.7$ for each sample. 
}
\label{fig:HI}
\end{figure}

The HI mass distributions in the two subsamples for galaxies with HI detection are presented in Fig.~\ref{fig:HI}. For each observation, solid line is the probability distribution of HI mass for galaxies with S/N$>4.7$, and dotted line is for galaxies with S/N$>6.5$. The probability is normalised to the total number of galaxies in each subsample. Galaxies observed by FAST have in general smaller HI masses. At low HI mass, more galaxies are detected by FAST due to the lower rms noise as discussed in Section 3.1. At high HI mass, galaxies observed by FAST still have lower HI mass compared with the ones in ALFALFA. In Section 3.1 for the 6 test galaxies that have been detected in both observations, we have seen that for the galaxies with proper measurements in both observations, the derived HI masses are quite consistent. Besides, the difference of the HI mass derived in the two observations due to different observation techniques is less than 0.03 dex. 
Therefore, there should exist intrinsic difference in HI mass of the galaxies in the two subsamples. For example, as seen in Fig.~\ref{fig:Mstar_z}, the galaxies observed by FAST is statistically a bit redder than the galaxies in ALFALFA, which is consistent with the fact that the FAST galaxies have in general lower HI mass. For ALFALFA galaxies with S/N$<6.5$, the HI masses may be over-estimated, because FAST observations of higher S/N indicates smaller HI masses for the test galaxies as seen in Fig.~\ref{fig:test_6galaxies}.

\begin{table}
    \caption{Number of galaxies with HI detection for the massive red spirals in the ALFALFA region and observed by FAST in this work. Results are given for signal-to-noise ratios greater than 6.5 and 4.7 respectively, and the corresponding detection rates are listed in parentheses.}
    \label{tab:samples}
    \begin{tabular}{lllr}
    \hline
     & All red spirals & S/N > 6.5  & S/N > 4.7\\
    \hline
    ALFALFA  & 166     & 60 (36.1$\%$)  &74 (44.6$\%$) \\[2pt]
    FAST     & 113     & 71 (62.8$\%$)  &75 (66.4$\%$) \\[2pt]
    \hline
    \end{tabular}
    \label{table:numbers}  
\end{table}

\begin{figure}
\hspace{-0.4cm}
\resizebox{8cm}{!}{\includegraphics{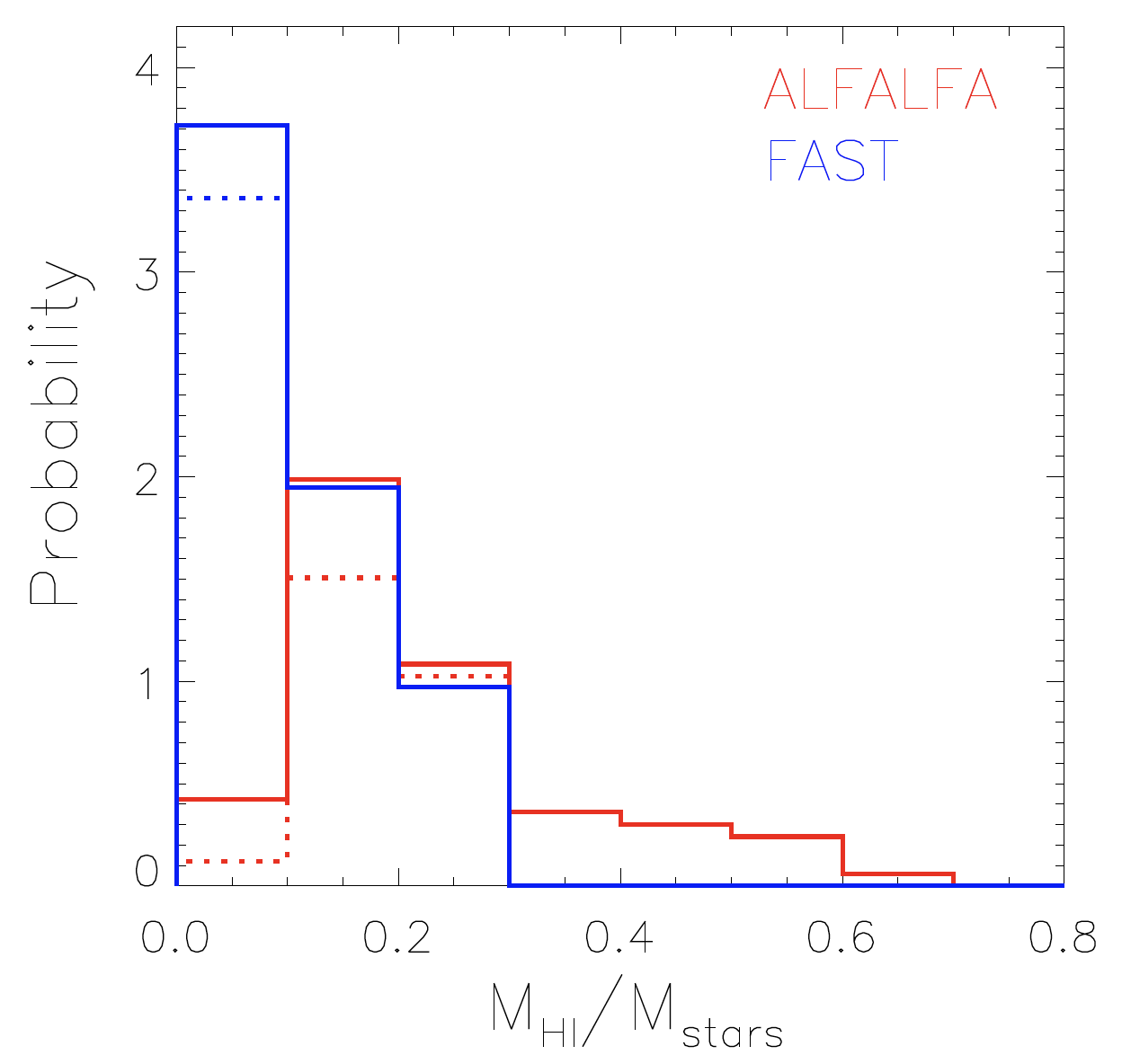}}\\
\caption{
Ratio of HI mass to stellar mass for the massive red spiral galaxies with HI detection, for ALFALFA (red) and FAST (blue) subsamples. The probability
is normalised to the total number of galaxies in each subsample. Solid histograms show probability distributions of the ratio for galaxies with $S/N>4.7$. Dotted histograms are the distributions of galaxies with $S/N>6.5$, which are on top of the solid histograms in the high ratio bins. 
}
\label{fig:HIratio}
\end{figure}

Fig.~\ref{fig:HIratio} shows the probability distributions of the HI mass to stellar mass ratio for the massive red spiral galaxies with HI detection in the two observations. Galaxies observed by FAST have in general lower HI-to-stellar mass ratio, due to the fact that more galaxies with low HI mass are detected, and the fact that galaxies observed by FAST have in general lower HI mass than those in ALFALFA as seen in Fig.~\ref{fig:HI}. The median ratio for galaxies with S/N>4.7 is 18.3 per cent in the ALFALFA sample, and is 8.5 per cent in the FAST sample. The latter value is even lower than the median ratio of HI-detected red elliptical galaxies observed in ALFALFA, which is 12 per cent as shown in Figure 14 of \citet{guorui2020}, again due to the fact that galaxies detected by FAST in this work reach lower HI mass than those observed in ALFALFA.
For galaxies covered by ALFALFA, there should be more galaxies with HI-to-stellar mass ratio less than around 10 per cent, but are not detected with the observational set of ALFALFA.

\section{color profile of massive red spirals}
The massive red spiral galaxies of \citet{guorui2020} used in this work are selected from the red sequence galaxies in the dust-corrected u-r color -- stellar mass diagram, with red color being the total color of the galaxies. While both the results from \citet{guorui2020} and from this work show that many of the red spirals contain prominent amount of HI gas, we investigate further where the HI mass comes from in these overall red galaxies. 

We examined the images of these massive red spirals from the DESI Legacy Imaging Surveys \citep{Legacy2019},
including Beijing-Arizona Sky Survey \citep{zouhu2017,zouhu2019}, Mayall z-band Sky Survey \citep{silva2016}, and DECam Legacy Survey \citep{blum2016}. These surveys provide optical imaging of g, r, and z bands over the sky area of about 14000 deg$^2$. The average depths are about 2--3 mag deeper than the SDSS. From the DESI imaging data, we found that many galaxies
have yellow and/or blue disks, although the bulge/central components of the galaxies are red without exception. A first look at the images of the galaxies with and without HI detection indicated that many of the galaxies with HI detection seem to have a yellow or blue disk, while the galaxies without HI detection tend to have orange/red disks. We therefore study in detail the color profiles of the massive red spirals, to see if there indeed exists a correlation between the HI mass and the color of the disk for these galaxies.

Cutouts of the massive red spirals are obtained from Data Release 8 of the DESI Legacy Imaging Surveys\footnote{\url{https://www.legacysurvey.org/dr8/}}, which include images of $g, r, z$ bands for each galaxy. Sigma-clipping is applied to both $g$ and $r$ band images to estimate the background and background noise. After subtracting the median background level from each pixel, galaxy images are analysed for pixels with flux above 3 times the background noise. 
Elliptical isophotes are fitted to the galaxy image of $r$ band, with python package $photutils.isophote$, using an iterative method as described by \citet{jedrzejewski1987}. Then the $g$ band image is sampled at the same geometry of elliptical isophotes derived from the $r$ band image, and color $g-r$ at each isophote is calculated as the difference of magnitudes in the two bands. When fitting elliptical isophotes for each galaxy, position angle and ellipticity of the isophotes are not required to be fixed, to represent better the properties of the outer disks. We have tested that with fixed position angle and ellipticity, the resulting statistics regarding color profiles shown in the following remain qualitatively similar.

\begin{figure*}
\hspace{-0.4cm}
\resizebox{16cm}{!}{\includegraphics{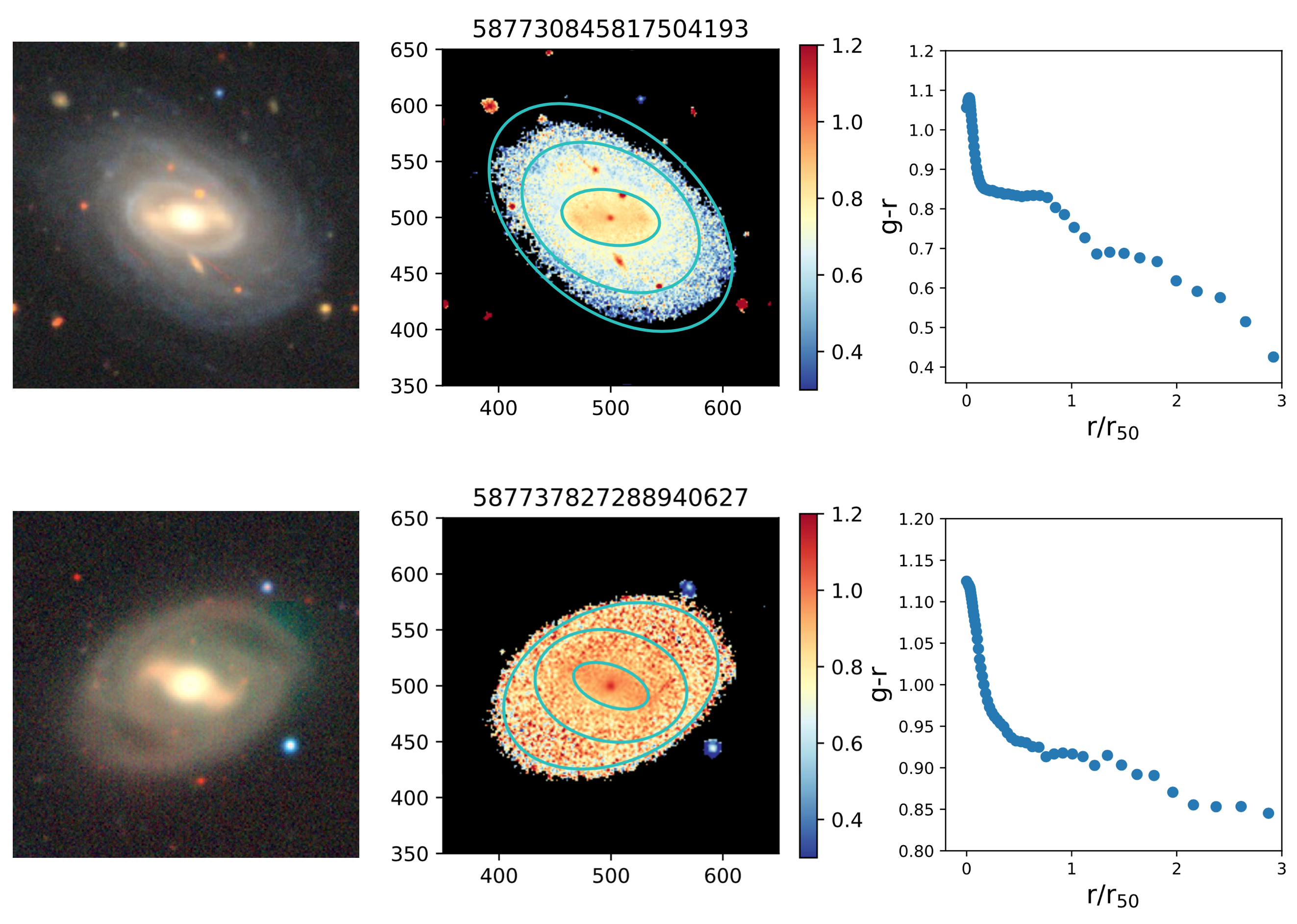}}\\
\caption{
Two example galaxies in the massive red spiral sample are shown, with the upper panels presenting a galaxy with blue outer disk, and the lower panels presenting a galaxy with all components red. For each galaxy, the left panel shows the $g, r, z$ bands colour image cutout from Data Release 8 of the DESI Legacy Imaging Surveys. The middle panel gives the $g-r$ color map of the galaxy, for pixels of flux above 3 times the background noise. On top of the color map, three circles are over-plotted representing elliptical isophotes with major axises equal to r$_{50}$, 2\,r$_{50}$ and 3\,r$_{50}$. The right panel shows color profile of the galaxy, with $g-r$ color at different elliptical isophotes as a function of the semimajor axis length of the isophotes. The semimajor axis lengths are normalized to r$_{50}$ of the galaxies in the x-axis.
}
\label{fig:example_color}
\end{figure*}

Fig.~\ref{fig:example_color} gives images and color profiles of two example galaxies in the massive red spiral sample. The left panels show their $g, r, z$-bands image cutouts from the DESI Legacy Imaging Surveys, and the middle panels show their $g-r$ color map. The example galaxy shown in the upper panels has a red central component but an obvious blue outer disk component, and its HI mass detected by FAST is $10^{10.03}M_{\odot}$. The galaxy shown in the lower panels has red components from center to edge, and has no HI detection by FAST. In the right panels, color profiles of the galaxies are shown, with $g-r$ color at different elliptical isophotes as a function of the semimajor axis length of the isophotes normalized to r$_{50}$ of the galaxy.  r$_{50}$ is the elliptical Petrosian half-light radius from the NASA-Sloan Atlas\footnote{\url{https://data.sdss.org/datamodel/files/ATLAS_DATA/ATLAS_MAJOR_VERSION/nsa.html}} \citep{Blanton2011}. Both galaxies have declining color profile from center to edge. The galaxy with HI detection in the upper panels has $g-r$ color less than 0.8 at radius greater than its r$_{50}$, and becomes less than 0.6 at radius greater than 2\,r$_{50}$. On the other hand, the galaxy without HI detection in the lower panels is always red in the color profile with $g-r$ color greater than 0.8.

\begin{figure*}
\hspace{-0.4cm}
\resizebox{18cm}{!}{\includegraphics{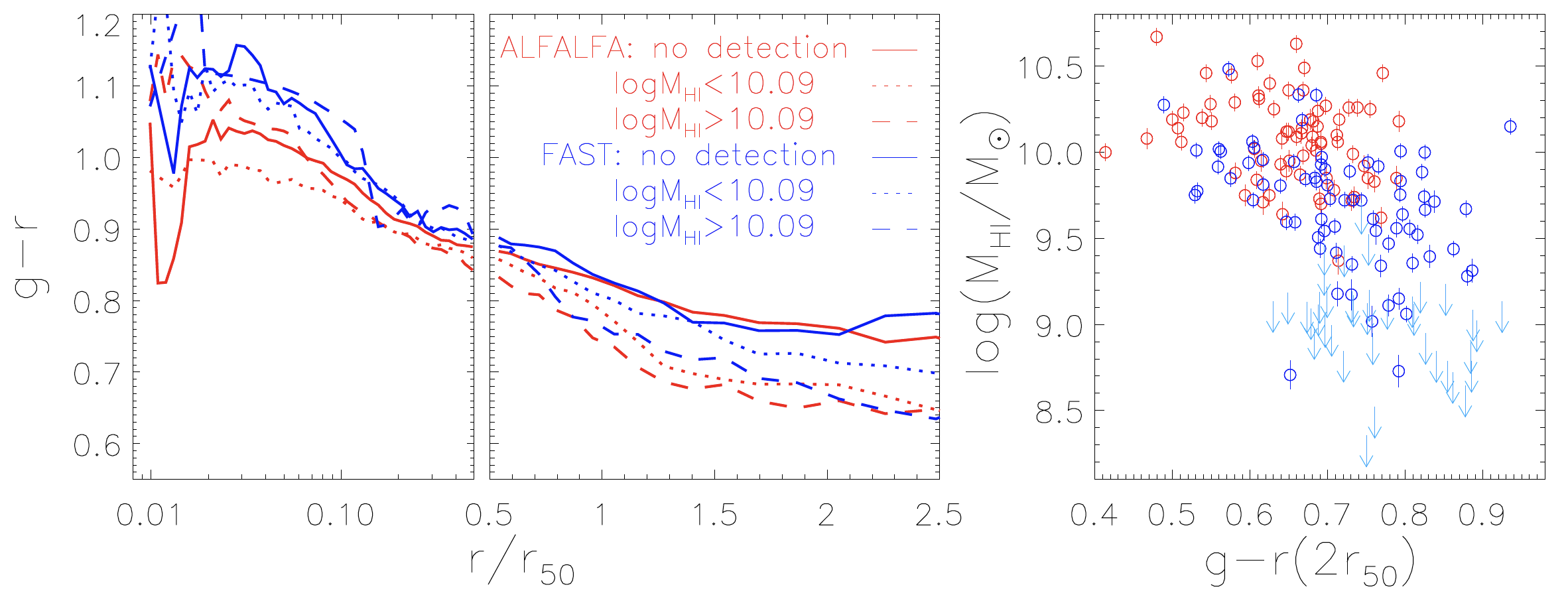}}
\caption{
Left and middle panels: color profile for the massive red spiral galaxies with and without HI detection in observations of ALFALFA (red) and FAST (blue). The results are presented in logarithmic (left panel) and in linear (middle panel) scale in x-axis, to show clearly the difference between subsamples at small radius and at large radius. Solid lines indicate the median color profile for galaxies without HI detection. Galaxies with HI detection are further divided into two subsamples by the median HI mass of galaxies detected in ALFALFA. The dotted lines present results for the subsample with lower HI mass, and dashed lines present results for the subsample with higher HI mass.  
Right panel: relation between galaxy HI mass and $g-r$ color at 2\,r$_{50}$ for each galaxy. Error bars indicate HI mass error estimation as described in section 2.2. For galaxies without HI detection in FAST observation, the tail ends of the light blue arrows correspond to the upper limit of their HI masses in the y-axis. For galaxies with no HI detection covered by ALFALFA, the median color at 2\,r$_{50}$ is 0.75, with 16 and 84 percentiles of 0.68 and 0.82. 
}
\label{fig:color_profile}
\end{figure*}

We analyse color profiles for all galaxies in the sample of massive red spirals. One galaxy is not in the region of DESI Legacy Imaging Surveys, and 5 galaxies have no r$_{50}$ provided in the NASA-Sloan Atlas catalog \citep{Blanton2011},
which leave us 273 massive red spirals in total. Fig.~\ref{fig:color_profile} shows the results of color profiles of these galaxies. In the left two panels, the median color profiles for galaxies with and without HI detection are plotted in different radius ranges. Solid red (blue) line is for ALFALFA (FAST) galaxies without HI detection (with S/N $<4.7$). Dotted and dashed red lines give median color profiles for galaxies with 50 per cent lower and higher HI mass in the subsample covered by ALFALFA, divided by their median HI mass of $10^{10.09}M_{\odot}$. For galaxies with HI detection observed by FAST, we also divide them into two subsamples with HI mass higher and lower than $10^{10.09}M_{\odot}$, to be compared directly with the subsamples of the ALFALFA galaxies.
Within radius of around 0.5\,r$_{50}$, the left panel shows that the galaxies observed by FAST are on average a bit redder than the galaxies in ALFALFA. This is consistent with the overall redder color of the FAST galaxies than the galaxies in ALFALFA as seen in Fig.~\ref{fig:Mstar_z}. Nevertheless, there is no obvious difference between subsamples with and without HI detection in both observations.
When looking at the color profile at larger radius as shown in the middle panel, deviations start to show up. Galaxies without HI detection are on average redder, especially at radius greater than 1.5\,r$_{50}$. For galaxies with HI detection, the ones with larger HI mass appear to be bluer than the ones with smaller HI mass, for galaxies both observed by FAST and in ALFALFA.

In the right panel of Fig.~\ref{fig:color_profile}, we show the trend seen in the left and middle panels in another way, by plotting the relation between galaxy color of outer disk (represented by the color at 2\,r$_{50}$) and galaxy HI mass detected in the two observations. Galaxies with HI detection (with S/N $>4.7$) in the ALFALFA and FAST observations are plotted by red and blue circles respectively.
For these galaxies in both observations, there exits a clear correlation that galaxies with larger HI mass are bluer in color at its outer radius. The HI mass detected should therefore mainly come from the outer blue disks, although in total the galaxies appear red due to the dominant red central component.
Note that in the right panel of Fig.~\ref{fig:color_profile}, the most right blue point represents a galaxy observed by FAST, which has relatively large HI mass of $10^{10.15}M_{\odot}$, but red color of 0.94 at 2\,r$_{50}$. We have checked that the disk is blue at even larger radius, and therefore is not conflict with the picture described above. 

For galaxies without HI detection (with S/N $<4.7$) in the FAST observation, we calculate their HI mass upper limit, and plot them using light blue arrows in the right panel of Fig.~\ref{fig:color_profile}, where the tail ends of the arrows correspond to their HI mass upper limits in the y-axis. For these galaxies, we use equation (2) to derive the S$_{21}$ upper limit by setting S/N=4.7. W$_{50}$ is adopted to be the median value for galaxies with HI detection. $\sigma_{\rm rms}$ is calculated on the two continuum regions around the potential HI signal of each galaxy as described in section 2.2. Then the HI mass upper limit is calculated according to equation (3). Their median g-r color at 2\,r$_{50}$ is 0.75 , with 16 and 84 percentiles to be 0.69 and 0.86. For galaxies covered by ALFALFA and have no HI detection, the median g-r color at 2 times of r$_{50}$ is 0.75, and the 16 and 84 percentiles are 0.68 and 0.82. On average, these galaxies are red at outer radius, with old stellar populations and do not have enough cold gas to be detected.

\section{Conclusions and discussions}

In this work we use the FAST radio telescope to observe HI content of 113 optically selected massive red spiral galaxies. 75 of the galaxies have HI detection with S/N greater than 4.7. Compared with the red spirals in the same sample selected by \citet{guorui2020} but have been observed in the ALFALFA survey, the observation by FAST in this work has lower rms noise and higher S/N. When applying the same limit of S/N, the detection rate of observation in FAST is higher, with more galaxies of low HI mass detected. For galaxies observed either in ALFALFA or by FAST, the large fraction of galaxies with HI detection is consistent with previous studies that many optically color/SFR selected red/quiescent spirals can contain large amount of atomic gas \citep{zhang2019, guorui2020}.

We also observe 6 massive red spirals that have been observed by ALFALFA, and compare in detail the spectra and HI mass derived in the two observations. For galaxies with spectra of S/N higher than 6.5 in ALFALFA, 
which can be considered as fairly safe detections \citep{saintonge2007}, the flux of the HI lines is consistent in the two observations for spectra not affected by RFI. The
HI mass from the two observations are consistent, not affected by the different integration time adopted in the two observations. For galaxies with spectra of lower signal-to-noise ratio, observation by FAST provides lower HI mass than that of ALFALFA, mainly due to the much larger line widths identified in ALFALFA. 

To study why many of the massive red spirals contain significant amount of atomic gas, we further analyse their optical images and study their color profiles in detail. We find that the massive red spirals selected according to the galaxy total color are not always all red from center to edge. Many of the red spirals have obvious blue disks in the outer region. There exists a clear correlation between HI mass and color of the outer disk of these galaxies, where galaxies with higher HI mass have in general bluer outer disks. 
These results suggest the failure of the global optical color to identify truly passive galaxies, and a close association between the cold gas content and the outer disc of galaxies. Both points have been found in previous works, as discussed below.

A decade ago, \citet{cortese2012} pointed out that at high stellar masses, optical colours are not sufficient to distinguish between star forming and quiescent galaxies. They found that optically selected red spirals lie on the UV-optical blue sequence, and are actually star-forming systems. The study of \citet{cortese2020} further demonstrated that claims of the existence of HI-rich, passive discs largely result from the failure to measure their outer star formation activities, and passive discs typically have more than 0.5 dex less HI than their active counterparts. In a recent work, \citet{zhoushuang2021} studied the massive red spiral sample of \citet{guorui2020} matched with MaNGA, and found that the optically selected red spirals are mostly green/blue in NUV-r, indicating on-going star formation in these galaxies. They found that the optically selected red spirals have on average young stellar populations in the outer region, consistent with our finding that many red spirals have blue outer disk and therefore high gas fraction.  

Galaxies are multi-component systems. Previous studies have suggested that cold gas is more tightly related/determined by single component of a galaxy, and independent of other components. For example, \citet{cook2019} found little dependence of the presence of a photometric bulge on the global HI gas content of local galaxies. \citet{namiki2021} showed that at fixed stellar mass and SFR, the presence of small-scale structures in galaxy images is linked to the total HI gas content of galaxies. \citet{wangjing2011} revealed that galaxies with higher HI fractions have bluer outer discs with more star-formation. Although we have not applied decomposition to the sample galaxies in this work or studied the relation between color of different galactic components and cold gas mass, our results also indicate that cold gas content is more related to the spatially resolved properties such as properties of disk component than the global properties of galaxies.

A missing point in this work is the identification of a direct connection between HI mass and star formation in the blue discs. Such a connection has been found at the outskirts of early-type galaxies \citep{yildz2017}. Furthermore, \citet{thilker2007} and \citet{lemonias2011} have shown that a significant fraction ($\sim 4\% - 30\%$) of both late and early type galaxies could have an extended UV (XUV) disk, which is an indication of recent star formation. Galaxies with an XUV-disk are nearly twice as gas-rich as their non-XUV-disk counterparts with a similar morphological type\citep{thilker2007}. These findings imply that a centrally quenched galaxy, given a large neutral gas reservoir, could also have recent star formation activities in the outskirt and thus a bluer outer disk. A large fraction of red spiral galaxies with HI detection in our sample could also be XUV-disk galaxies. We would defer a further study using UV photometry in a future paper.

\section*{Acknowledgements}
We thank the reviewer for useful comments and constructive suggestions that helped improve the manuscript.
We thank Cheng Li, Yougang Wang, Jing Wang and Marko Krco for helpful discussions and comments.
This work was supported by the National Key Program for Science and Technology Research Development of China 2018YFE0202902, the National Natural Science Foundation of China (NSFC) under grants 11988101, 11733002, and K.C.Wong Education Foundation.
ZZ is supported by NSFC grants No. U1931110 and 12041302, and CAS Interdisciplinary Innovation Team (JCTD-2019-05).
RL acknowledges the support of NSFC (Nos 11773032,12022306), and the science research grants from the China Manned Space Project (Nos CMS-CSST-2021-B01,CMS-CSST-2021-A01).
H. Zou acknowledges the support of the NSFC (grant No. 12120101003) and the supports of the China Manned Space Project (grant Nos. CMS-CSST-2021-A02 and CMS-CSST-2021-A04).

Five-hundred-meter Aperture Spherical radio Telescope (FAST) is a National Major Scientific Project built by the Chinese Academy of Sciences. Funding for the project has been provided by the National Development and Reform Commission. FAST is operated and managed by the National Astronomical Observatories, Chinese Academy of Sciences.

The Legacy Surveys consist of three individual and complementary projects: the Dark Energy Camera Legacy Survey (DECaLS; Proposal ID \#2014B-0404; PIs: David Schlegel and Arjun Dey), the Beijing-Arizona Sky Survey (BASS; NOAO Prop. ID \#2015A-0801; PIs: Zhou Xu and Xiaohui Fan), and the Mayall z-band Legacy Survey (MzLS; Prop. ID \#2016A-0453; PI: Arjun Dey). DECaLS, BASS and MzLS together include data obtained, respectively, at the Blanco telescope, Cerro Tololo Inter-American Observatory, NSF’s NOIRLab; the Bok telescope, Steward Observatory, University of Arizona; and the Mayall telescope, Kitt Peak National Observatory, NOIRLab. The Legacy Surveys project is honored to be permitted to conduct astronomical research on Iolkam Du’ag (Kitt Peak), a mountain with particular significance to the Tohono O’odham Nation.

\section*{Data Availability}
Table \ref{table:catalog} that provides the results of the 113 massive red spirals observed by FAST is available online.
The data from ALFALFA survey is available in the paper of \citet{haynes2018}. The images of galaxies are available from the DESI Legacy Imaging Surveys at: \url{https://www.legacysurvey.org/dr8/}.
Other data produced in this article will be shared on reasonable request to the corresponding author.


\bibliographystyle{mnras}
\bibliography{red_spiral}



\bsp	
\label{lastpage}
\end{document}